\documentclass[twocolumn,showpacs,aps,prb,superscriptaddress,floatfix,amsmath,amssymb]{revtex4}

\usepackage{graphicx}
\usepackage{dcolumn}
\usepackage{bm}
\usepackage{subfigure}

\begin{document}

\title{Magnons and a two-component spin gap in $\mathrm{FeV_{2}O_{4}}$}

\author{G. J. MacDougall}
\email{gmacdoug@illinois.edu}
\affiliation{Department of Physics and the Seitz Materials Research Laboratory, University of Illinois at Urbana-Champaign, Urbana, Illinois, USA, 61801}

\author{I. Brodsky}
\affiliation{Department of Physics and the Seitz Materials Research Laboratory, University of Illinois at Urbana-Champaign, Urbana, Illinois, USA, 61801}

\author{A. A. Aczel}
\affiliation{Quantum Condensed Matter Division, Oak Ridge National Laboratory, Oak Ridge, Tennessee, 37831, USA}

\author{V. O. Garlea}
\affiliation{Quantum Condensed Matter Division, Oak Ridge National Laboratory, Oak Ridge, Tennessee, 37831, USA}

\author{G. E. Granroth}
\affiliation{Quantum Condensed Matter Division, Oak Ridge National Laboratory, Oak Ridge, Tennessee, 37831, USA}
\affiliation{Neutron Data Analysis and Visualization Division , Oak Ridge National Laboratory, Oak Ridge, Tennessee 37831, USA}

\author{A. D. Christianson}
\affiliation{Quantum Condensed Matter Division, Oak Ridge National Laboratory, Oak Ridge, Tennessee, 37831, USA}

\author{T. Hong}
\affiliation{Quantum Condensed Matter Division, Oak Ridge National Laboratory, Oak Ridge, Tennessee, 37831, USA}

\author{H. D. Zhou}
\affiliation{Department of Physics and Astronomy, University of Tennessee, Knoxville, TN, 37996, USA}

\author{S. E. Nagler}
\affiliation{Quantum Condensed Matter Division, Oak Ridge National Laboratory, Oak Ridge, Tennessee, 37831, USA}
\affiliation{CIRE, University of Tennessee, Knoxville, TN, 37996, USA}

\date{\today}
\begin{abstract}

The spinel vanadates have become a model family for exploring orbital order on the frustrated pyrochlore lattice, and recent debate has focused on the symmetry of local crystal fields at the cation sites. Here, we present neutron scattering measurements of the magnetic excitation spectrum in $\mathrm{FeV_2O_4}$, a recent example of a ferrimagnetic spinel vanadate which is available in single crystal form. We report the existence of two emergent magnon modes at low temperatures, which draw strong parallels with the closely related material, $\mathrm{MnV_2O_4}$. We were able to reproduce the essential elements of both the magnetic ordering pattern and the dispersion of the inelastic modes with semiclassical spin wave calculations, using a minimal model that implies a sizeable single-ion anisotropy on the vanadium sublattice. Taking into account the direction of ordered spins, we associate this anisotropy with the large trigonal distortion of $VO_6$ octahedra, previously observed via neutron powder diffraction measurements. We further report on the spin gap, which is an order-of-magnitude larger than that observed in $\mathrm{MnV_2O_4}$. By looking at the overall temperature dependence, we were able to show that the gap magnitude is largely associated with the ferro-orbital order known to exist on the iron sublattice, but the contribution to the gap from the vanadium sublattice is in fact comparable to what is reported in the Mn compound. This reinforces the conclusion that the spin canting transition is associated with the ordering of vanadium orbitals in this system, and closer analysis indicates closely related physics underlying orbital transitions in $\mathrm{FeV_2O_4}$ and $\mathrm{MnV_2O_4}$.

\end{abstract}


\maketitle

\section{Introduction}



The spinel vanadates ($\mathrm{AV_2O_4}$) are an important model family for the study of orbital order, and notable for the frustrated pyrochlore network of spin and orbitally-active vanadium cations therein\cite{lee10}. The major outstanding question in the study of these compounds is the nature of the orbital ordered state at low temperatures, and scattering studies are playing an important role in determining the relative importance of sub-dominate interaction and crystal field terms in the magnetic Hamiltonian. Here, we present inelastic neutron scattering data on $\mathrm{FeV_2O_4}$ which contribute to this important conversation. $\mathrm{FeV_2O_4}$ is a ferrimagnetic spinel, characterized by two orbitally active cation sites, and shown by previous diffraction studies to have three structural and two magnetic transitions\cite{katsufuji08,macdougall12,nii12}. We report on the existence of two inelastic spin-wave modes in the low temperature magnetic phases whose dispersions are well described by a model that assumes significant trigonal crystal fields at the vanadium site. Our data further show that the spin gap has a temperature dependence that reflects both magnetic transitions, and suggests that a full description of magnetism in this material requires consideration of the orbital order which exists on both cation sites.

Spinel vanadates with divalent transition metal atoms on the A-site are Mott insulators, and the octahedrally coordinated vanadium ($V^{3+}$) cations on the pyrochlore sublattice have S=1 spin and orbital triplet degrees-of-freedom in the ideal cubic phase. Nearly without exception, each exhibits a cubic-to-tetragonal structural transition with decreasing temperature, which partially lifts the orbital degeneracy and alters magnetic properties. In materials with non-magnetic A-site cations ($A^{2+} \in (Zn^{2+}, Cd^{2+}, Mg^{2+})$), the cubic-tetragonal transition precedes the onset of $\mathbf{Q}= 2\pi (0,0,1)$ antiferromagnetic order at lower temperature\cite{lee10,wheeler10}. The ferrimagnetic materials $\mathrm{MnV_2O_4}$  and $\mathrm{FeV_2O_4}$ are observed to have successive collinear Neel and canted antiferromagnetic transitions\cite{adachi05,suzuki07,garlea08,chung08,macdougall12}, where the spin canting is coincident with the onset of the low-temperature tetragonal structure. Though generally associated with orbital order, the nature of these structural transitions and the exact configuration of electron orbitals at low temperatures are issues of active interest.

Discussions of orbital order have largely focused on the relative importance of two proposed patterns: a form of antiferro-orbital order (AFO) containing an alternating pattern of real orbitals in the $ab$-plane \cite{tsunetsugu03,motome04}, and a form of ferro-orbital order (FOO) where each site is occupied by a complex superposition of orbitals\cite{tchernyshyov04}. The two patterns are associated with slightly different tetragonal space groups, with the FOO pattern containing an additional glide plane symmetry (space group $I4_1/amd$, rather than $I4_1/a$), but both were derived assuming a tetragonally distorted cubic crystal field environment. Recent developments, however, have indicated that a full understanding of the orbital ground state might require a proper treatment of the \textit{trigonal} crystal fields which exist in the spinel structure when the fractional coordinate, $x$, deviates from its ideal value of 0.25\cite{yaresko2008}. Trigonal fields have been invoked to explain neutron scattering data in $\mathrm{MgV_2O_4}$\cite{wheeler10} and the existence of a high field transition in $\mathrm{CdV_2O_4}$\cite{mun14}. First-principles calculations have indicated that trigonal crystal fields play a defining role for the low-temperature states of $\mathrm{MnV_2O_4}$\cite{sarkar09}  and $\mathrm{FeV_2O_4}$\cite{sarkar11}. The ``quantum $120^{\circ}$ model'' of Chern $\textit{et al.}$\cite{chern10} was able to explain 
the experimental picture surrounding $\mathrm{MnV_2O_4}$ by assuming that the trigonal crystal fields were a \textit{dominant} contribution to the magnetic Hamiltonian. Our own neutron powder diffraction (NPD) work subsequently demonstrated that the observed magnetic ground state of $\mathrm{FeV_2O_4}$ is consistent with the predictions of the quantum $120^{\circ}$ model in the strong spin-orbit coupling limit\cite{macdougall12}.

The current study follows up on our initial neutron powder diffraction study of $\mathrm{FeV_2O_4}$, and presents inelastic neutron scattering data on single crystals.  $\mathrm{FeV_2O_4}$ is quite unique among the spinel vanadates, in that it has an orbital doublet degree-of-freedom on the A-site $Fe^{2+}$ cation, in addition to the orbitally active vanadium cation on the spinel `B'-site. It is observed to have three distinct structural transitions, evolving from high-temperature cubic (HTC) to high-temperature tetragonal (HTT) to face-centered orthorhombic (FCO) to higher symmetry low-temperature tetragonal (LTT) structure with decreasing temperature\cite{katsufuji08,macdougall12,nii12}. The HTT-FCO and FCO-LTT structural transitions are further associated with the onset of collinear and canted spin structures, respectively, involving both of the cation sites\cite{macdougall12,kang12}. The lowest temperature canted phase has additionally been shown to exhibit a net ferroelectric moment, which can be manipulated with moderate applied magnetic field\cite{zhang12,kismarahardja13}. The physics underlying these transitions has been argued mostly from analogy to other spinel systems. The highest temperature structural transition is thought to be the result of a Jahn-Teller transition and the onset of ferro-orbital order on the $Fe^{2+}$ site, similar to $\mathrm{FeCr_2O_4}$\cite{shirane64,goodenough64,bordacs09,tsuda10} and consistent with the distortion of the local $FeO_4$ tetrahedra\cite{katsufuji08,macdougall12,nii12}. The HTT-FCO transition is thought to be driven by antiferromagnetic exchange between the two cation sites, consistent with spin-only chromates\cite{bordacs09} and $\mathrm{MnV_2O_4}$\cite{garlea08}. In analogy to $\mathrm{MnV_2O_4}$, we have also argued that the lowest transition can be understood to result from the onset of orbital order on the vanadium sublattice\cite{macdougall12} and pointed to the predictions of the quantum 120$^\circ$ model\cite{chern10}.

In this article, we further this discussion by presenting neutron scattering measurements of spin-wave spectra in a large single crystal of $\mathrm{FeV_2O_4}$. Two low energy modes are identified, with strong parallels to observed modes in $\mathrm{MnV_2O_4}$\cite{garlea08,chung08}, but fit to a model which more appropriately includes a local $< 111 >$ single-ion anisotropy to encompass trigonal fields on the vanadium sites. Using these fits, we argue that both [001] anisotropy on the iron site and the $<111>$ anisotropy on the vanadium sites are  essential for a proper description of the low-temperature physics. Implications for the low temperature orbital order and the parallels to $\mathrm{MnV_2O_4}$ are discussed.

\section{Experimental Methods}

Single crystals of $\mathrm{FeV_2O_4}$ were grown by the float zone method, as described elsewhere\cite{macdougall12}. Crystals were characterized first by heat capacity, using a Quantum Design Physical Property Measurement System at the National High Magnetic Field Laboratory in Tallahassee. The resulting data were previously published in the Supplementary Materials of Ref.~\onlinecite{macdougall12}, and are presented again in Figure~\ref{fig:characterization}. Bulk magnetization was measured using a Vibrating Sample Magnetometer at the University of Illinois at Urbana-Champaign, with applied fields of H = 500Oe along the cubic (001) direction, and plotted with the heat capacity data for direct comparison.

Neutron scattering experiments were performed using instruments at both the Spallation Neutron Source (SNS) and the High Flux Isotope Reactor (HFIR) at Oak Ridge National Laboratory. Spin waves were first measured using the SEQUOIA Fine Resolution Chopper Spectrometer at the SNS\cite{granroth2006,granroth2010}. The majority of measurements used $E_i$=55meV neutrons and the coarse chopper on SEQUOIA\cite{granroth2010}, giving energy resolution of approximately 3 meV at the elastic line. One 3g crystal was mounted in a closed cycle refrigerator with the pseudo-cubic (HK0) plane horizontal, and spectra at three different temperatures (T = 4K, 85K and 120K) were built from measurements taken at 0.5$^\circ$ steps over a 100$^\circ$ range using the Mantid framework\cite{taylor2012mantid}. Background scattering from the cryostat and sample can were measured separately and subtracted from the data. 
Plots of SEQUOIA data were made using the Horace software package\cite{horace}.

Further measurements to explore the temperature dependence of the spin gap were performed at the HFIR, using the HB3 and CTAX triple-axis (TA) spectrometers. Both sets of measurements made use of the same crystal explored with SEQUOIA, oriented in the same scattering plane. For the lowest temperatures, the large energy transfers involved demanded the use of thermal neutrons. The HB3 spectrometer was employed, with a PG002 monochromator and analyzer, 48$^{\prime}$-40$^{\prime}$-40$^{\prime}$-120$^{\prime}$ collimation and $E_f$=14.7meV neutrons. Higher order contamination was removed with two PG filters. Finer resolution measurements were performed at temperatures near the upper magnetic transitions, using the CTAX cold TA spectrometer, with guide-open-80$^{\prime}$-open collimation. The energy of the scattered neutrons was fixed at $E_f$ = 5 meV. Higher order contamination was removed by a cooled Be filter placed between the sample and analyzer.

\section{Results and Discussion}

\begin{figure}[t]
\begin{center}
\includegraphics[width=\columnwidth]{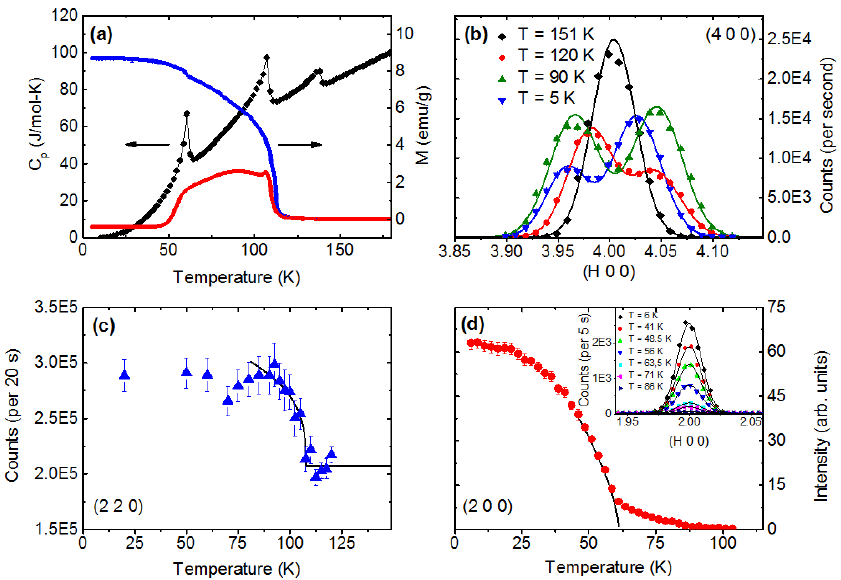}
\caption{\textbf{(a)}Heat capacity and magnetization data from the single crystal sample from the current neutron scattering study. Peaks in heat capacity identify transitions at 138 K, 107 K, and 60 K. Magnetization was measured under field-cooled (upper/blue line) and zero-field-cooled (lower/red line) conditions, with field oriented along the cubic (001) direction. \textbf{(b)} Variation of the structural (400) Bragg peak, as measured by elastic neutron scattering. Temperatures were chosen in each of the previously identified phases. Solid lines are fits to Gaussian curves. \textbf{(c)} Elastic scattering intensity of the cubic (220) position, reflecting the square magnitude of ordered moments on the $Fe^{2+}$ sublattice and \textbf{(d)} (200), reflecting the canting component of spins on the $V^{3+}$ sublattice.  Data in panels (c) and (d) were measured using the HB3 and CTAX triple-axis spectrometers, respectively. Solid lines represent fits to Eq.~\ref{eq:transition} and are discussed in the main text. }\label{fig:characterization}
\end{center}
\end{figure}

Characterization and elastic neutron scattering data shown in Fig.~\ref{fig:characterization} largely confirm the temperature and nature of the phase transitions reported by our previous NPD work\cite{macdougall12}. Peaks in the heat capacity data (Fig.~\ref{fig:characterization} (a)) identify phase transitions at 138 K, 107 K and 60 K, and can be immediately associated with previously identified Jahn-Teller ($T_c$), the collinear N$\mathrm{\acute{e}}$el ($T_{N1}$) and the spin canting ($T_{N2}$) transitions, respectively. Magnetization versus temperature data for the current single crystal sample are plotted in the same panel, measured in both field-cooled (solid line) and zero-field-cooled (dashed line) configurations. As with measurements on powders\cite{macdougall12,zhang12}, the field-cooled curve reveals an increase in net magnetization at the 107 K and 60 K transitions, reflecting the onset of collinear and canted ferrimagnetism. The divergence of ZFC and FC lines below $T_{N1}$ can be associated with ferrimagnetic domain formation.

Fig.~\ref{fig:characterization}(b) shows scans of elastic neutron scattering intensity across the cubic (400) Bragg position. Temperatures are representative of the four distinct structural phases, identified above, and the peak profiles exhibit an evolution in-line with the known cubic-HTT-FCO-LTT sequence of transitions. The single cubic peak at T = 200 K splits at 120 K, reflecting the cubic-tetragonal transition at $T_{c}$ = 140 K. The scan at T = 90 K shows a shift in scattering intensity from the high-angle to low angle peak, with the latter peak broader than resolution and best described by two Gaussians, as expected for the FCO phase. The T = 5 K scan is again described by two peaks, with intensity distributed between them with opposite sense to what is observed at T = 120 K. As with our NPD study\cite{macdougall12}, and confirmed by a subsequent single-crystal x-ray study\cite{nii12}, there is no indication of further structural transitions below $T_{N2}$ = 60 K.

\begin{figure}[t]
\begin{center}
\includegraphics[width=\columnwidth]{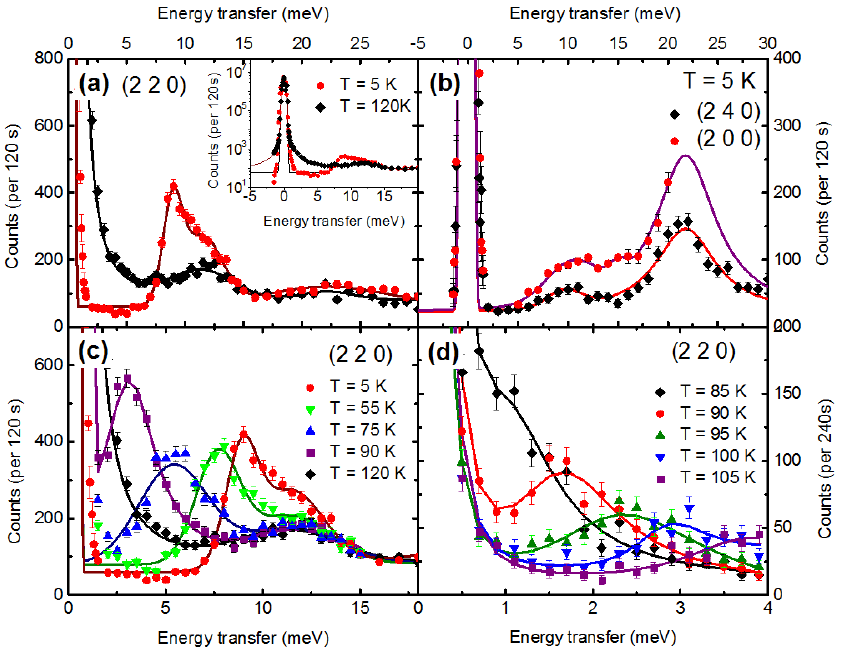}
\caption{\textbf{(a)} Constant-Q scans on the HB3 triple-axis spectrometer at the cubic (2 2 0) position. Shown are scans at temperatures above and well below the ordering temperatures in $\mathrm{FeV_2O_4}$ on logarithmic (inset) and linear (main panel) scales.  \textbf{(b)} Constant-Q scans at the (2 0 0) and (2 4 0) positions, which are symmetrically equivalent. \textbf{(c)}\textbf{(d)} show the temperature evolution of the constant-Q scans at (2 2 0), as measured with the HB3 and CTAX spectrometers, respectively. }\label{fig:TA_inelastic}
\end{center}
\end{figure}

The existence of a magnetic transition at $T_{N1}$ = 110K is confirmed by the emergence of spin wave excitations and the temperature dependence of the spin gap, as laid out below. Fig.~\ref{fig:characterization}(c) further shows temperature dependence of elastic scattering at the (220) Bragg position, as determined by scans of neutron energy with constant-\textbf{Q} (e.g. Inset, Fig.~\ref{fig:TA_inelastic}(a)). The intensity of this peak reflects the ordered moment on the iron sublattice, and serves as an order parameter for the Neel antiferromagnetic state. The intensity of the Bragg peak at the (200) position was measured via radial scans in the elastic channel, and is plotted in  Fig.~\ref{fig:characterization}(d). The (200) Bragg peak reflects a breaking of a local glide plane symmetry preserved in the collinear spin state, and its intensity acts as an order parameter for spin canting. Data in Fig.~\ref{fig:characterization} (c) and (d) were fitted to the function

\begin{equation}
I(T) = I_0*(1-\frac{T}{T_N})^{2\beta} + const,
\label{eq:transition}
\end{equation}

in the temperature range $\frac{T}{T_N} > 0.75$, to extract approximate values for critical temperatures and exponents. The extracted critical exponents, $\beta_1 = 0.16 \pm 0.06$ and $\beta_2 = 0.32 \pm 0.03$, are broadly consistent with Ising transitions in two and three dimensions, respectively, though more detailed measurements would be required to comment further. The fitted transition temperatures, $T_{N1} = 107.5 K \pm 0.1 K$ and $T_{N2} = 61.1 K \pm 0.5 K$, in-line with our earlier estimates from heat capacity and NPD. A new feature revealed by the single-crystal measurements is the significant elastic intensity about the (200) position which is seen to persist to temperatures well above $T_{N2}$, and decrease monotonically with warming. In this temperature region, the (200) peaks are broader in \textbf{Q} than instrument resolution and can be associated with the existence of short-range spin canting correlations within the collinear Neel state. It is important to note that there is no change in the (400) Bragg intensity over the same temperature range, and so this effect cannot be simply associated with scattering from this structural peak with imperfectly filtered $\lambda/2$ neutrons. Spin canting is symmetrically allowed in the Fddd spacegroup of the FCO phase, and the observed short range correlations above the ordering transition is similar to what is seen in other frustrated geometries.

\begin{figure}[t]
\begin{center}
\includegraphics[width=\columnwidth]{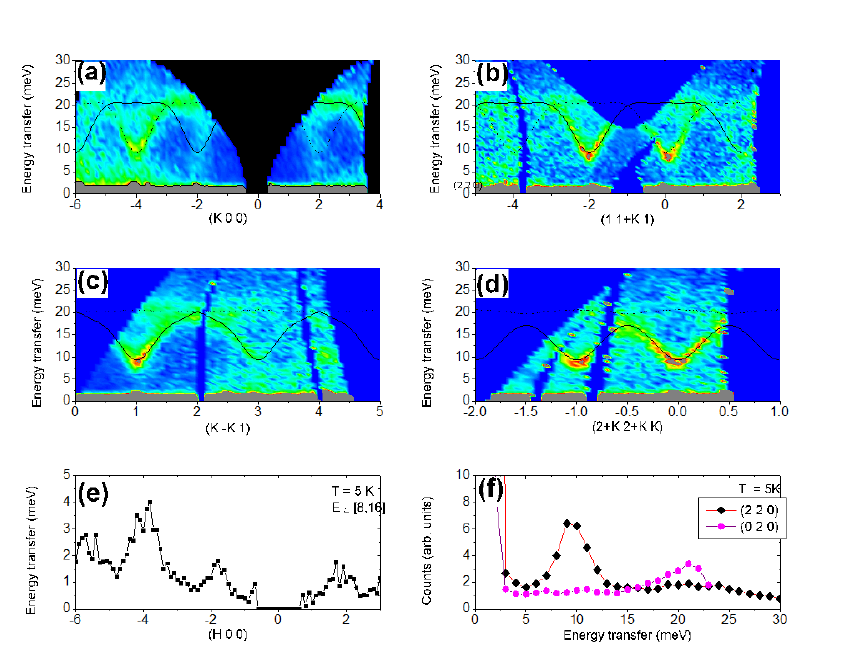}
\caption{\textbf{(a)-(d)} Contour plots of neutron scattering intensity in several energy-momentum transfer planes, showing the Q-dependence of magnon excitations along several different symmetric directions in reciprocal space. Especially notable are the dispersive magnetic excitations which arise about points (2 0 0), (4 0 0) and symmetric equivalents. \textbf{(e)} A plot of scattering intensity versus (H 0 0) in the energy range E = 8-16 meV, revealing peaks at (-4 0 0) and ($\pm$2 0 0). \textbf{(f)} Plot of scattering intensity versus energy transfer at (2 2 0) and (2 0 0). In all plots, we use the high-temperature cubic basis to index the reciprocal lattice cell.}\label{fig:SEQ_slices}
\end{center}
\end{figure}

\begin{figure}[t]
\begin{center}
\includegraphics[width=\columnwidth]{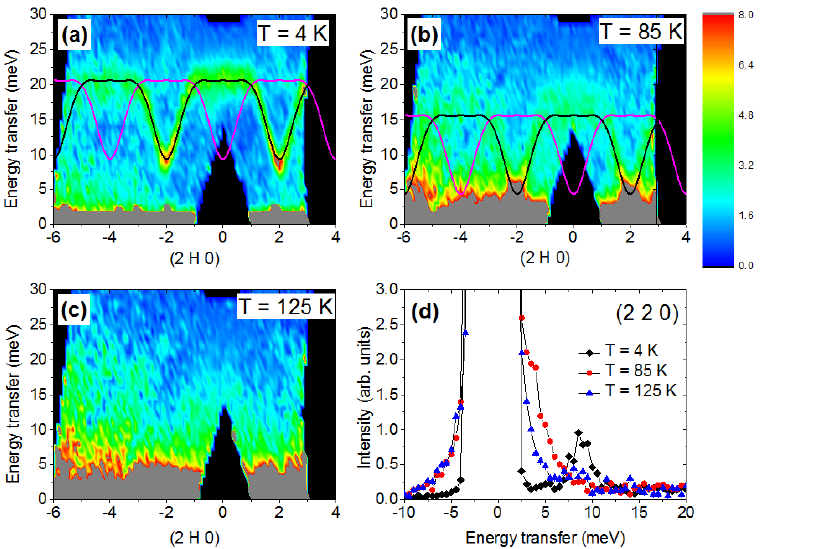}
\caption{\textbf{(a)-(c)} Plots of inelastic neutron scattering intensity along the line (2 H 0) in reciprocal space, as measured with the SEQUOIA spectrometer. Shown are data taken at T = 4 K (a), 85 K (b) and 125 K (c). Solid lines are spin-wave fits described in the main text. Curves in (b) are shifted by 5 meV. \textbf{(d)} Plots of neutron scattering intensity versus energy transfer for all three temperatures at the point (2 2 0).}\label{fig:SEQ_temp}
\end{center}
\end{figure}

Main inelastic results from the SEQUOIA chopper spectrometer are summarized in Figures~\ref{fig:SEQ_slices} and \ref{fig:SEQ_temp}. At the lowest temperatures, we observe two distinct dispersive modes with energies below 25 meV, and interpret them as magnons of the ferrimagnetic ordered state. The dispersion of these modes is plotted along five different symmetry directions in Fig.~\ref{fig:SEQ_slices}(a)-(d) and Fig.~\ref{fig:SEQ_temp}(a). The dominant mode is roughly twenty times as intense as the ``weak'' mode, and has shape and bandwidth reminiscent of the lowest energy mode reported by Chung \textit{et al.} for $\mathrm{MnV_2O_4}$\cite{chung08}. The minima coincide with the zone-centers of the diamond sublattice, consistent with the previous claim that this excitation is associated with the motion of A-site spins. The second mode is more difficult to discern in the contour plots of Fig.~\ref{fig:SEQ_slices}, but plots of neutron scattering intensity versus (H00) (Fig.~\ref{fig:SEQ_slices}(e)) and energy transfer (Fig.~\ref{fig:SEQ_slices}(f)) indicate that the weak mode has minima (maxima) where the strong mode has maxima (minima), with an equally large spin gap. These statements are confirmed below via triple-axis measurements.

A full theoretical construction to describe magnetic excitations in $\mathrm{FeV_2O_4}$ must take into account the local ionic levels of both $Fe^{2+}$ and $V^{3+}$ cations\cite{buyers1971}. Here, we show instead that the essential elements of the observed spin excitations at low temperatures are captured by a minimal spin-wave model, with the inclusion of appropriate single-ion anisotropy terms.

Dispersion data were fit using semiclassical spin-wave calculations, assuming an ideal cubic structure and the Hamiltonian
\[\mathcal{H} = \sum\limits_{< i,j >} J_{i,j} \mathbf{S_i} \cdot \mathbf{S_j} +  \sum\limits_i D_i (\mathbf{S_i} \cdot \mathbf{\hat{n}_i})^2\]
and the results are shown as solid lines in Fig.~\ref{fig:SEQ_slices}. Here, the sums are over all spins, including both cation sites, and interactions are truncated beyond nearest neighbors for the $Fe^{2+}$ sites (nearest neighbor Fe-V interactions) and next-nearest neighbors for the $V^{3+}$ sites (nearest neighbor V-V and nearest neighbor Fe-V interactions). The single-ion anisotropies are constrained to be along the (001) direction for the iron site, consistent with conclusions from x-ray diffraction and magnetization\cite{katsufuji08}, and along the local $<111>$ directions for the vanadium sites, consistent with an important role for the trigonal crystal field. The exchange parameter, $J_{i,j}$ is also fitted differently when describing V-V pairs in ($J_{BB}$) or out of ($J^{'}_{BB}$) the tetragonal a-b plane.

The canting angle of spins in the ground state was determined self-consistently through fits of the inelastic data, and found to be within error equal to the value inferred from NPD\cite{macdougall12}. Further fit parameters are listed in Table~1. The spin-wave model explains the dispersion of the two observed modes in all directions, and identifies them as having a majority contribution from the motion of iron cations. The model predicts the existence of four additional modes with primarily vanadium character, not discernable in our inelastic neutron scattering data below energy transfers of 35 meV. One possible explanation for their absence is that they exist at energies above our measurement range; as shown in the Supplementary Information\cite{SM_reference}, spin-wave curves associated with the parameters in Table~1 indeed predict the four remaining modes reside in the energy range 30 - 60 meV. However, we further note that the ordered moment on the vanadium site is roughly 5 times weaker than the order iron moment\cite{macdougall12}, and thus the missing modes are expected to be much less intense than those reported here.

\begin{table}[h]
\begin{tabular}{|c|c|c|c|c|c|}
	\hline
$J_{AB}$ & $J_{BB}$ & $J^{'}_{BB}$ & $D^{[001]}_{A}$ & $D^{<111>}_{B}$ & canting angle \\
    \hline
2.9(1)  & 15(2)  & -0.6(9)   & -0.01(4) & -9.5(5) & 55(2) \\ \hline

\end{tabular}
\caption{ Exchange and anisotropy parameters inferred from fits of TOF neutron scattering data collected at T = 5 K to a semiclassical spin-wave model. }
\end{table}

Of the parameters listed in Table~1, the exchange parameter $J_{AB}$ is most tightly constrained by the available data, but there was significant freedom in chosing $J_{BB}$ and $J^{'}_{BB}$. The given values provide a good description of the current data, but further constraints on $J_{BB}$ and $J^{'}_{BB}$ will likely require measurements of the four remaining modes. It is interesting to note, however, that even a good description of all dispersion curves was possible only if we allowed the parameter $J^{'}_{BB}$, describing the exchange between V-V nearest neighbor pairs in the (101) direction, to be ferromagnetic. An antiferromagnetic $J_{BB}$ and ferromagnetic $J^{'}_{BB}$ are in fact perfectly consistent with the 2-in-2-out spin structure stabilized in  $\mathrm{FeV_2O_4}$. Importantly, both anisotropy parameters $D^{[001]}_{A}$ and $D^{<111>}_{B}$ are predicted to contribute to the spin gap. The best fit value for $D^{<111>}_{B}$ is large and negative. A large, negative anisotropy favoring spins in the $<111>$ directions is consistent with the assumptions underlying the quantum 120$^{\circ}$ model of Chern \textit{et al.}\cite{chern10}

In Fig.~\ref{fig:SEQ_temp}, we show a representative contour plot of the low-energy excitation spectrum at three temperatures, T = 5 K, 85 K and 125 K, corresponding to measurements in the canted spin, the collinear ferrimagnetic, and the paramagnetic states, respectively. One can see that the overall shape and bandwidth of the dominant mode is unchanged as the system evolves from the canted to the collinear spin state, with only a loss of scattering intensity resulting from a smaller ordered moment size at 85 K. This is as expected, as the spins on the iron sublattice are unaffected at the canting transition. Above $T_{N1}$, the modes vanish entirely, confirming they are magnetic in origin.  The most notable change with temperature is the magnitude of the spin gap, which is $\Delta \sim 9 meV$ in the LTT phase but barely distinguishable from zero in Fig.~\ref{fig:SEQ_temp}(d) at T = 85 K.

To explore the temperature dependence of the magnetic states more fully, SEQUOIA results were supplemented by targeted triple-axis measurements over the range 5 K $<$ T $<$ 120 K. Figure~\ref{fig:TA_inelastic}(a) shows a comparison of scattering intensity at the (220) reciprocal lattice position versus energy transfer well-below and above the upper magnetic transition at $T_{N1}$ = 110K.  The choice of (220) was dictated by the minimum of the dominant band in Fig.~\ref{fig:SEQ_slices}. The inset is plotted on a logarithmic scale, and reveals significant quasi-elastic scattering intensity at the higher temperature which collapses into the gapped magnon modes below the ordering transitions. The modes themselves are shown more clearly on a linear scale  in the main panel. Similar base temperature scans are shown in Figure~\ref{fig:TA_inelastic}(b) at the (200) position- the location of the lowest-Q minimum of the weak mode in Fig.~\ref{fig:SEQ_slices}- and at (240)- a symmetrically equivalent position with larger dynamic range on the TA spectrometer.

Scans in both panels indicate multiple magnetic excitations below 25 meV. The dominant and weak magnon modes identified with SEQUOIA are confirmed in Fig.~\ref{fig:TA_inelastic}(a), where the greater intensity branch has a gap 8.9 meV $\pm$ 0.1 meV and the lower intensity branch appears much broader and is located around 23 meV. Both peaks are also present at (240) with the relative intensities reversed. In addition, the scans seem to indicate that an excitation exists with E$\approx$12 meV. An equivalent excitation is seen at all measured \textbf{Q}, with an intensity that decreases with increasing scattering angle.  As shown in Fig.~\ref{fig:TA_inelastic}(c), this peak also persists to temperatures as high as 120 K. The origin of this peak is a subject for future investigation.

\begin{figure}[t]
\begin{center}
\includegraphics[width=\columnwidth]{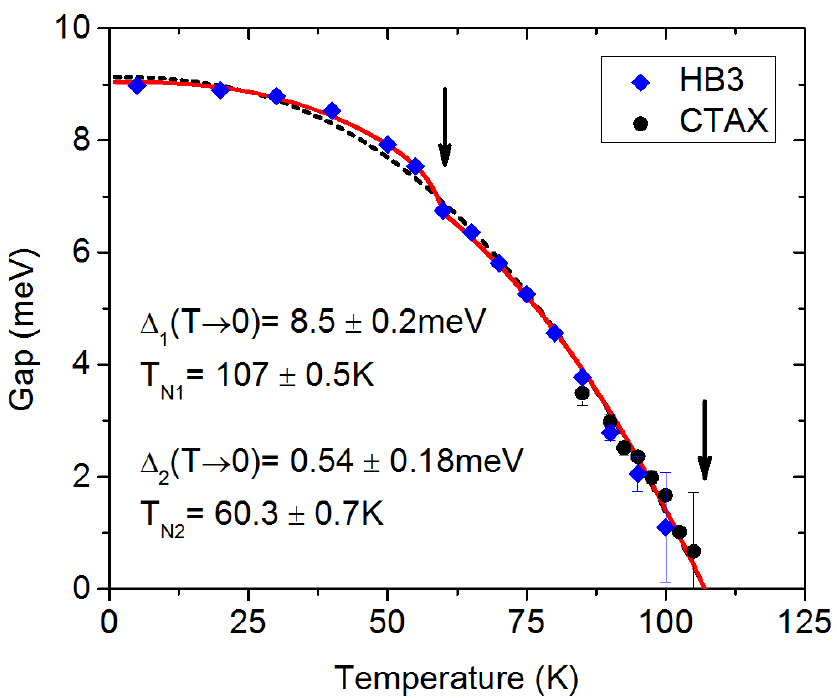}
\caption{ Temperature dependence of the excitation gap at the cubic (2 2 0) position in $\mathrm{FeV_2O_4}$. Included are data from the HB3 (diamonds) and CTAX (circles) spectrometers. Lines are fits to power-law temperature dependence, assuming one (dashed) or two (solid) order-parameters. The latter was a better description of the data, and the parameters from that fit are shown explicitly.}\label{fig:OP}
\end{center}
\end{figure}

To track the temperature dependence of the spin gap, we took scans at the (220) position with limited energy transfer and temperature increasing by 5 K increments from 5 K to 120 K. Select scans taken using a thermal TA instrument are shown in Fig.~\ref{fig:TA_inelastic}(c), and using a cold TA in Fig.~\ref{fig:TA_inelastic}(d). Solid lines represent the results of fitting to a series of Lorentzian peaks convolved with instrument resolution, with the energy of the 12 meV excitation held constant below T = 100 K to stabilize the fitting. The fitted gap is plotted as a function of temperature in Fig.~\ref{fig:OP}. Results from the two datasets are consistent in the temperature region overlap, and the overall gap function is seen to rise up in a smooth, mean-field like way from the upper N$\mathrm{\acute{e}}$el temperature.

The temperature dependence was characterized using both single and double order-parameter functions, with the best fits shown as dashed and solid lines, respectively. The single OP fit  gives $T_{N1}$ = 107.7 K $\pm$ 0.5 K, within error equal to other measurements of critical temperature. However, a far superior description of the data is provided by the two-OP fit, which yields $T_{N1}$ = 107.0 K $\pm$ 0.5 K and $T_{N2}$ = 60.3 K $\pm$ 0.7 K. These values independently confirm the critical temperatures obtained from heat capacity, NPD and elastic neutron measurements. The increase of the gap at the canting transition, $T_{N2}$, is reminiscent of the opening of a gap from zero in $\mathrm{MnV_2O_4}$ and reinforces the notion that this temperature can be associated with the ordering of orbitals on the vanadium sublattice. It is natural to ascribe the second, higher-temperature contribution to the gap function to the ferro-orbital order that exists on the iron sublattice below T = 140 K. Thus, the two-OP gap function presented in Fig.~\ref{fig:OP} indicates that orbital order on both cations sites is playing a defining role in this material.

The most distinguishing feature of the inelastic spectrum of $\mathrm{FeV_2O_4}$ is the order-of-magnitude larger spin-gap over other spinel vanadates. However, the data in Fig.~\ref{fig:OP} imply that this is primarily due to the orbital order on the A-site, combined with spin-orbit coupling and the large ordered iron moment, and has little to do with the physics of vanadium cations. Other distinguishing features of $\mathrm{FeV_2O_4}$, including the (0 0 1) easy-axis observed via magnetization and associated magnetostrictive effects observed with x-ray diffraction\cite{katsufuji08}, can also be explained by a strong single ion anisotropy for the spin on the A-site, as made clear by first principles calculations\cite{sarkar11}.

When focusing instead on the contribution to the gap from the vanadium sublattice, the magnitude and temperature dependence is comparable to what is reported for $\mathrm{MnV_2O_4}$\cite{garlea08}. We have noted in the past\cite{macdougall12} that the elastic ordering pattern of spins in $\mathrm{FeV_2O_4}$ at low temperatures is consistent with the predictions of the quantum 120$^\circ$ model\cite{chern10} of Chern \textit{et al.} in the strong spin-orbit coupling limit. In this model, developed to explain observations on $\mathrm{MnV_2O_4}$, the in-plane direction of vanadium spins in the canted state is set by a competition between orbital exchange interactions and coupling to local trigonal distortions, which prefer spins to point along primary cubic axes and local $<11>$ spin directions, respectively. Intuitively then, the statement that $\mathrm{FeV_2O_4}$ lies in the strong spin-orbit coupling limit is equivalent to highlighting the dominant role of trigonal distortions, consistent with the success of our spin-wave model above. In the paper of Chern, it was concluded that $\mathrm{MnV_2O_4}$ lies in the opposite, strong orbital exchange limit. This is surprising in the current context, as $\mathrm{MnV_2O_4}$ has both a larger trigonal distortion than $\mathrm{FeV_2O_4}$ and a larger spin-gap arising from the vanadium sublattice. A possible resolution comes from noting that the classification of $\mathrm{MnV_2O_4}$ as being in the strong-exchange limit was based primarily on the directions for vanadium spins stated in the NPD paper of Garlea \textit{et al.}\cite{garlea08}. However, the unpolarized NPD data of Garlea \textit{et al.} were unable to uniquely determine in-plane spin directions\cite{shirane59}, and so the authors were careful to say that the decision to constrain the \textit{in-plane} portion of the ordered spins to point along primary axes was an assumption of the structure refinement\cite{garlea08}.
As a further point of interest, one recent neutron diffraction study on single crystals of $\mathrm{MnV_2O_4}$\cite{magee_thesis} largely confirms the structure of Garlea \textit{et al.}, but reports a slightly smaller canting angle of 54.9$^\circ$ for that material, nearly identical to the present case. Thus, the spin and orbital structures of $\mathrm{MnV_2O_4}$ and $\mathrm{FeV_2O_4}$ may be more closely related than previously thought.

A strong counter-argument to this conclusion is the different tetragonal space groups reported for $\mathrm{MnV_2O_4}$ and $\mathrm{FeV_2O_4}$. In fact, the $I4_1/amd$ space group reported for $\mathrm{FeV_2O_4}$\cite{katsufuji08,macdougall12} contains a glide-plane symmetry that is expected to be broken in the quantum 120$^\circ$ model, and normal mode analysis of recent x-ray diffraction data suggests fundamentally different orbitally ordered states for the Mn and Fe compounds\cite{nii12}. The trigonal distortion in $\mathrm{FeV_2O_4}$ is large, but smaller than in $\mathrm{MnV_2O_4}$. We suggest that the underlying assumption of a dominant (i.e. infinite) trigonal distortion may be violated in the case of $\mathrm{FeV_2O_4}$. Detailed measurements of the inelastic spectrum of the two compounds, and an analysis which incorporates the full orbital and spin degrees-of-freedom in the  $t_{2g}$ manifold\cite{buyers1971} may be key to a complete understanding of these systems.

\section{Summary and Conclusions}

In conclusion, we have examined single crystal samples of $\mathrm{FeV_2O_4}$ with elastic and inelastic neutron scattering. The sequence of structural and magnetic phase transitions identified in powders are confirmed by heat capacity, magnetization and elastic neutron scattering on crystals. Inelastic neutron scattering measurements reveal two magnon branches below the ferrimagnetic transition at 110 K, and a semiclassical spin wave analysis was shown to be successful if single ion anisotropies are assumed to be in the (0 0 1) direction for the iron sublattice and the local $<111>$ directions on the vanadium sublattice. The spin-gap was measured to be 8.9 $\pm$ 0.1 meV, mostly resulting from the ferro-orbital order on the A-site sublattice. The small increase in the gap from the vanadium sublattice is comparable to other materials, and consistent with the identification of the canting transition at 60 K with an orbital ordering transition on the B-site. Similarities in the measured inelastic spectra and the local direction of ordered spins suggest that the physics describing $\mathrm{MnV_2O_4}$ and $\mathrm{FeV_2O_4}$ systems are closely related. Reconciling the different crystallographic space groups in the two materials may require future measurements of the inelastic spectrum.

\section{Acknowledgements}

 Authors would like to acknowledge valuable discussions with S. Hahn and R. Fishman at ORNL. Research at the Spallation Neutron Source and the High Flux Isotope Reactor was sponsored by the U. S. Department of Energy, Office of Basic Energy Sciences, Scientific User Facilities Division. G.J.M and I.B. are further supported by U.S. Department of Energy, Office of Basic Energy Sciences, Division of Materials Sciences and Engineering under Award DE-FG02-07ER46453.  H.D.Z. thanks the support of the JDRD program of the University of Tennessee.


\begin{thebibliography}{31}
\expandafter\ifx\csname natexlab\endcsname\relax\def\natexlab#1{#1}\fi
\expandafter\ifx\csname bibnamefont\endcsname\relax
  \def\bibnamefont#1{#1}\fi
\expandafter\ifx\csname bibfnamefont\endcsname\relax
  \def\bibfnamefont#1{#1}\fi
\expandafter\ifx\csname citenamefont\endcsname\relax
  \def\citenamefont#1{#1}\fi
\expandafter\ifx\csname url\endcsname\relax
  \def\url#1{\texttt{#1}}\fi
\expandafter\ifx\csname urlprefix\endcsname\relax\def\urlprefix{URL }\fi
\providecommand{\bibinfo}[2]{#2}
\providecommand{\eprint}[2][]{\url{#2}}

\bibitem[{\citenamefont{Lee et~al.}(2010)}]{lee10}
\bibinfo{author}{\bibfnamefont{S.-H.} \bibnamefont{Lee}} \bibnamefont{et~al.},
  \bibinfo{journal}{J.\ Phys.\ Soc.\ Jap.} \textbf{\bibinfo{volume}{79}},
  \bibinfo{pages}{011004} (\bibinfo{year}{2010}).

\bibitem[{\citenamefont{Katsufuji et~al.}(2008)\citenamefont{Katsufuji, Suzuki,
  Takei, Shingu, Kato, Osaka, Takata, Sagayama, and Arima}}]{katsufuji08}
\bibinfo{author}{\bibfnamefont{T.}~\bibnamefont{Katsufuji}},
  \bibinfo{author}{\bibfnamefont{T.}~\bibnamefont{Suzuki}},
  \bibinfo{author}{\bibfnamefont{H.}~\bibnamefont{Takei}},
  \bibinfo{author}{\bibfnamefont{M.}~\bibnamefont{Shingu}},
  \bibinfo{author}{\bibfnamefont{K.}~\bibnamefont{Kato}},
  \bibinfo{author}{\bibfnamefont{K.}~\bibnamefont{Osaka}},
  \bibinfo{author}{\bibfnamefont{M.}~\bibnamefont{Takata}},
  \bibinfo{author}{\bibfnamefont{H.}~\bibnamefont{Sagayama}}, \bibnamefont{and}
  \bibinfo{author}{\bibfnamefont{T.}~\bibnamefont{Arima}},
  \bibinfo{journal}{J.\ Phys.\ Soc.\ Japan} \textbf{\bibinfo{volume}{77}},
  \bibinfo{pages}{053708} (\bibinfo{year}{2008}).

\bibitem[{\citenamefont{MacDougall et~al.}(2012)\citenamefont{MacDougall,
  Garlea, Aczel, Zhou, and Nagler}}]{macdougall12}
\bibinfo{author}{\bibfnamefont{G.~J.} \bibnamefont{MacDougall}},
  \bibinfo{author}{\bibfnamefont{V.~O.} \bibnamefont{Garlea}},
  \bibinfo{author}{\bibfnamefont{A.~A.} \bibnamefont{Aczel}},
  \bibinfo{author}{\bibfnamefont{H.~D.} \bibnamefont{Zhou}}, \bibnamefont{and}
  \bibinfo{author}{\bibfnamefont{S.~E.} \bibnamefont{Nagler}},
  \bibinfo{journal}{Phys.\ Rev.\ B} \textbf{\bibinfo{volume}{86}},
  \bibinfo{pages}{060414(R)} (\bibinfo{year}{2012}).

\bibitem[{\citenamefont{Nii et~al.}(2012)\citenamefont{Nii, Sagayama, Arima,
  Aoyagi, Sakai, Maki, Nishibori, Sawa, Sugimoto, Ohsumi et~al.}}]{nii12}
\bibinfo{author}{\bibfnamefont{Y.}~\bibnamefont{Nii}},
  \bibinfo{author}{\bibfnamefont{H.}~\bibnamefont{Sagayama}},
  \bibinfo{author}{\bibfnamefont{T.}~\bibnamefont{Arima}},
  \bibinfo{author}{\bibfnamefont{S.}~\bibnamefont{Aoyagi}},
  \bibinfo{author}{\bibfnamefont{R.}~\bibnamefont{Sakai}},
  \bibinfo{author}{\bibfnamefont{S.}~\bibnamefont{Maki}},
  \bibinfo{author}{\bibfnamefont{E.}~\bibnamefont{Nishibori}},
  \bibinfo{author}{\bibfnamefont{H.}~\bibnamefont{Sawa}},
  \bibinfo{author}{\bibfnamefont{K.}~\bibnamefont{Sugimoto}},
  \bibinfo{author}{\bibfnamefont{H.}~\bibnamefont{Ohsumi}},
  \bibnamefont{et~al.}, \bibinfo{journal}{Phys.\ Rev.\ B}
  \textbf{\bibinfo{volume}{86}}, \bibinfo{pages}{125142}
  (\bibinfo{year}{2012}).

\bibitem[{\citenamefont{Wheeler et~al.}(2010)\citenamefont{Wheeler, Lake,
  Islam, Reehuis, Steffens, Guidi, and Hill}}]{wheeler10}
\bibinfo{author}{\bibfnamefont{E.~M.}~\bibnamefont{Wheeler}},
  \bibinfo{author}{\bibfnamefont{B.}~\bibnamefont{Lake}},
  \bibinfo{author}{\bibfnamefont{A.~T.~M.~N.}~ \bibnamefont{Islam}},
  \bibinfo{author}{\bibfnamefont{M.}~\bibnamefont{Reehuis}},
  \bibinfo{author}{\bibfnamefont{P.}~\bibnamefont{Steffens}},
  \bibinfo{author}{\bibfnamefont{T.}~\bibnamefont{Guidi}}, \bibnamefont{and}
  \bibinfo{author}{\bibfnamefont{A.~H.} \bibnamefont{Hill}},
  \bibinfo{journal}{Phys.\ Rev.\ B} \textbf{\bibinfo{volume}{82}},
  \bibinfo{pages}{140406(R)} (\bibinfo{year}{2010}).

\bibitem[{\citenamefont{Adachi et~al.}(2005)\citenamefont{Adachi, Suzuki, Kato,
  Osaka, Takata, and Katsufuji}}]{adachi05}
\bibinfo{author}{\bibfnamefont{K.}~\bibnamefont{Adachi}},
  \bibinfo{author}{\bibfnamefont{T.}~\bibnamefont{Suzuki}},
  \bibinfo{author}{\bibfnamefont{K.}~\bibnamefont{Kato}},
  \bibinfo{author}{\bibfnamefont{K.}~\bibnamefont{Osaka}},
  \bibinfo{author}{\bibfnamefont{M.}~\bibnamefont{Takata}}, \bibnamefont{and}
  \bibinfo{author}{\bibfnamefont{T.}~\bibnamefont{Katsufuji}},
  \bibinfo{journal}{Phys.\ Rev.\ Lett.} \textbf{\bibinfo{volume}{95}},
  \bibinfo{pages}{197202} (\bibinfo{year}{2005}).

\bibitem[{\citenamefont{Suzuki et~al.}(2007)\citenamefont{Suzuki, Nagai,
  Nohara, and Takagi}}]{suzuki07}
\bibinfo{author}{\bibfnamefont{T.}~\bibnamefont{Suzuki}},
  \bibinfo{author}{\bibfnamefont{H.}~\bibnamefont{Nagai}},
  \bibinfo{author}{\bibfnamefont{M.}~\bibnamefont{Nohara}}, \bibnamefont{and}
  \bibinfo{author}{\bibfnamefont{H.}~\bibnamefont{Takagi}},
  \bibinfo{journal}{J.\ Phys.:\ Condens.\ Matter}
  \textbf{\bibinfo{volume}{19}}, \bibinfo{pages}{144265}
  (\bibinfo{year}{2007}).

\bibitem[{\citenamefont{Garlea et~al.}(2008)}]{garlea08}
\bibinfo{author}{\bibfnamefont{V.~O.} \bibnamefont{Garlea}}
  \bibnamefont{et~al.}, \bibinfo{journal}{Phys.\ Rev.\ Lett.}
  \textbf{\bibinfo{volume}{100}}, \bibinfo{pages}{066404}
  (\bibinfo{year}{2008}).

\bibitem[{\citenamefont{Chung et~al.}(2008)\citenamefont{Chung, Kim, Lee, Sato,
  Suzuki, Katsumura, and Katsufuji}}]{chung08}
\bibinfo{author}{\bibfnamefont{J.~H.} \bibnamefont{Chung}},
  \bibinfo{author}{\bibfnamefont{J.~H.} \bibnamefont{Kim}},
  \bibinfo{author}{\bibfnamefont{S.~H.} \bibnamefont{Lee}},
  \bibinfo{author}{\bibfnamefont{T.~J.} \bibnamefont{Sato}},
  \bibinfo{author}{\bibfnamefont{T.}~\bibnamefont{Suzuki}},
  \bibinfo{author}{\bibfnamefont{M.}~\bibnamefont{Katsumura}},
  \bibnamefont{and}
  \bibinfo{author}{\bibfnamefont{T.}~\bibnamefont{Katsufuji}},
  \bibinfo{journal}{Phys.\ Rev.\ B} \textbf{\bibinfo{volume}{77}},
  \bibinfo{pages}{054412} (\bibinfo{year}{2008}).

\bibitem[{\citenamefont{Tsunetsugu and Motome}(2003)}]{tsunetsugu03}
\bibinfo{author}{\bibfnamefont{H.}~\bibnamefont{Tsunetsugu}} \bibnamefont{and}
  \bibinfo{author}{\bibfnamefont{Y.}~\bibnamefont{Motome}},
 \bibinfo{journal}{Phys.\ Rev.\ B} \textbf{\bibinfo{volume}{68}},
  \bibinfo{pages}{060405(R)} (\bibinfo{year}{2003}).


\bibitem[{\citenamefont{Motome and Tsunetsugu}(2004)}]{motome04}
\bibinfo{author}{\bibfnamefont{Y.}~\bibnamefont{Motome}} \bibnamefont{and}
  \bibinfo{author}{\bibfnamefont{H.}~\bibnamefont{Tsunetsugu}},
  \bibinfo{journal}{Phys.\ Rev.\ B} \textbf{\bibinfo{volume}{70}},
  \bibinfo{pages}{184427} (\bibinfo{year}{2004}).


\bibitem[{\citenamefont{Tchernyshyov}(2004)}]{tchernyshyov04}
\bibinfo{author}{\bibfnamefont{O.}~\bibnamefont{Tchernyshyov}},
  \bibinfo{journal}{Phys.\ Rev.\ Lett.} \textbf{\bibinfo{volume}{93}},
  \bibinfo{pages}{157206} (\bibinfo{year}{2004}).

\bibitem[{\citenamefont{Yaresko}(2008)}]{yaresko2008}
\bibinfo{author}{\bibfnamefont{A.~N.} \bibnamefont{Yaresko}},
  \bibinfo{journal}{Phys.\ Rev.\ B} \textbf{\bibinfo{volume}{77}},
  \bibinfo{pages}{115106} (\bibinfo{year}{2008}).

\bibitem[{\citenamefont{Mun et~al.}(2014)\citenamefont{Mun, Chern, Pardo,
  Rivadulla, Zapf, and Batista}}]{mun14}
\bibinfo{author}{\bibfnamefont{E.~D.} \bibnamefont{Mun}},
  \bibinfo{author}{\bibfnamefont{G.-W.} \bibnamefont{Chern}},
  \bibinfo{author}{\bibfnamefont{V.}~\bibnamefont{Pardo}},
  \bibinfo{author}{\bibfnamefont{F.}~\bibnamefont{Rivadulla}},
    \bibinfo{author}{\bibfnamefont{R.}~\bibnamefont{Sinclair}},
    \bibinfo{author}{\bibfnamefont{H.~D.}~\bibnamefont{Zhou}},
  \bibinfo{author}{\bibfnamefont{V.~S.} \bibnamefont{Zapf}}, \bibnamefont{and}
  \bibinfo{author}{\bibfnamefont{C.~D.} \bibnamefont{Batista}},
  \bibinfo{journal}{Phys.\ Rev.\ Lett.} \textbf{\bibinfo{volume}{112}},
  \bibinfo{pages}{017207} (\bibinfo{year}{2014}).

\bibitem[{\citenamefont{Sarkar et~al.}(2009)\citenamefont{Sarkar, Maitra,
  Valenti, and Saha-Dasgupta}}]{sarkar09}
\bibinfo{author}{\bibfnamefont{S.}~\bibnamefont{Sarkar}},
  \bibinfo{author}{\bibfnamefont{T.}~\bibnamefont{Maitra}},
  \bibinfo{author}{\bibfnamefont{R.}~\bibnamefont{Valenti}}, \bibnamefont{and}
  \bibinfo{author}{\bibfnamefont{T.}~\bibnamefont{Saha-Dasgupta}},
  \bibinfo{journal}{Phys.\ Rev.\ Lett.} \textbf{\bibinfo{volume}{102}},
  \bibinfo{pages}{216405} (\bibinfo{year}{2009}).

\bibitem[{\citenamefont{Sarkar and Saha-Dasgupta}(2011)}]{sarkar11}
\bibinfo{author}{\bibfnamefont{S.}~\bibnamefont{Sarkar}} \bibnamefont{and}
  \bibinfo{author}{\bibfnamefont{T.}~\bibnamefont{Saha-Dasgupta}},
  \bibinfo{journal}{Phys.\ Rev.\ B} \textbf{\bibinfo{volume}{84}},
  \bibinfo{pages}{235112} (\bibinfo{year}{2011}).

\bibitem[{\citenamefont{Chern et~al.}(2010)\citenamefont{Chern, Perkins, and
  Hao}}]{chern10}
\bibinfo{author}{\bibfnamefont{G.~W.} \bibnamefont{Chern}},
  \bibinfo{author}{\bibfnamefont{N.}~\bibnamefont{Perkins}}, \bibnamefont{and}
  \bibinfo{author}{\bibfnamefont{Z.}~\bibnamefont{Hao}},
  \bibinfo{journal}{Phys.\ Rev.\ B} \textbf{\bibinfo{volume}{81}},
  \bibinfo{pages}{125127} (\bibinfo{year}{2010}).

\bibitem[{\citenamefont{Kang et~al.}(2012)\citenamefont{Kang, Hwang, Kim, Lee,
  Kim, Kim, Kwon, Lee, Kim, Ueno et~al.}}]{kang12}
\bibinfo{author}{\bibfnamefont{J.-S.} \bibnamefont{Kang}},
  \bibinfo{author}{\bibfnamefont{J.}~\bibnamefont{Hwang}},
  \bibinfo{author}{\bibfnamefont{D.~H.} \bibnamefont{Kim}},
  \bibinfo{author}{\bibfnamefont{E.}~\bibnamefont{Lee}},
  \bibinfo{author}{\bibfnamefont{W.~C.} \bibnamefont{Kim}},
  \bibinfo{author}{\bibfnamefont{C.~S.} \bibnamefont{Kim}},
  \bibinfo{author}{\bibfnamefont{S.}~\bibnamefont{Kwon}},
  \bibinfo{author}{\bibfnamefont{S.}~\bibnamefont{Lee}},
  \bibinfo{author}{\bibfnamefont{J.-Y.} \bibnamefont{Kim}},
  \bibinfo{author}{\bibfnamefont{T.}~\bibnamefont{Ueno}}, \bibnamefont{et~al.},
  \bibinfo{journal}{Phys.\ Rev.\ B} \textbf{\bibinfo{volume}{85}},
  \bibinfo{pages}{165136} (\bibinfo{year}{2012}).

\bibitem[{\citenamefont{Zhang et~al.}(2012)\citenamefont{Zhang, Singh, Guillou,
  Simon, Breard, Caignaert, and Hardy}}]{zhang12}
\bibinfo{author}{\bibfnamefont{Q.}~\bibnamefont{Zhang}},
  \bibinfo{author}{\bibfnamefont{K.}~\bibnamefont{Singh}},
  \bibinfo{author}{\bibfnamefont{F.}~\bibnamefont{Guillou}},
  \bibinfo{author}{\bibfnamefont{C.}~\bibnamefont{Simon}},
  \bibinfo{author}{\bibfnamefont{Y.}~\bibnamefont{Breard}},
  \bibinfo{author}{\bibfnamefont{V.}~\bibnamefont{Caignaert}},
  \bibnamefont{and} \bibinfo{author}{\bibfnamefont{V.}~\bibnamefont{Hardy}},
  \bibinfo{journal}{Phys.\ Rev.\ B} \textbf{\bibinfo{volume}{85}},
  \bibinfo{pages}{054405} (\bibinfo{year}{2012}).

\bibitem[{\citenamefont{Kismarahardja et~al.}(2013)\citenamefont{Kismarahardja,
  Brooks, Zhou, Choi, Matsubayashi, and Uwatoko}}]{kismarahardja13}
\bibinfo{author}{\bibfnamefont{A.}~\bibnamefont{Kismarahardja}},
  \bibinfo{author}{\bibfnamefont{J.~S.} \bibnamefont{Brooks}},
  \bibinfo{author}{\bibfnamefont{H.~D.} \bibnamefont{Zhou}},
  \bibinfo{author}{\bibfnamefont{E.~S.} \bibnamefont{Choi}},
  \bibinfo{author}{\bibfnamefont{K.}~\bibnamefont{Matsubayashi}},
  \bibnamefont{and} \bibinfo{author}{\bibfnamefont{Y.}~\bibnamefont{Uwatoko}},
  \bibinfo{journal}{Phys.\ Rev.\ B} \textbf{\bibinfo{volume}{87}},
  \bibinfo{pages}{054432} (\bibinfo{year}{2013}).

\bibitem[{\citenamefont{Shirane et~al.}(1964)\citenamefont{Shirane, Cox, and
  Pickart}}]{shirane64}
\bibinfo{author}{\bibfnamefont{G.}~\bibnamefont{Shirane}},
  \bibinfo{author}{\bibfnamefont{D.~E.} \bibnamefont{Cox}}, \bibnamefont{and}
  \bibinfo{author}{\bibfnamefont{S.~J.} \bibnamefont{Pickart}},
  \bibinfo{journal}{J.\ Appl.\ Phys.} \textbf{\bibinfo{volume}{35}},
  \bibinfo{pages}{954} (\bibinfo{year}{1964}).

\bibitem[{\citenamefont{Goodenough}(1964)}]{goodenough64}
\bibinfo{author}{\bibfnamefont{J.~B.} \bibnamefont{Goodenough}},
  \bibinfo{journal}{J.\ Phys.\ Chem.\ Solids} \textbf{\bibinfo{volume}{25}},
  \bibinfo{pages}{151} (\bibinfo{year}{1964}).

\bibitem[{\citenamefont{Bordacs et~al.}(2009)\citenamefont{Bordacs, Varjas,
  Kezsmarki, Mihaly, Baldaddarre, Abouelsayed, Kuntscher, Ohgushi, and
  Tokura}}]{bordacs09}
\bibinfo{author}{\bibfnamefont{S.}~\bibnamefont{Bordacs}},
  \bibinfo{author}{\bibfnamefont{D.}~\bibnamefont{Varjas}},
  \bibinfo{author}{\bibfnamefont{I.}~\bibnamefont{Kezsmarki}},
  \bibinfo{author}{\bibfnamefont{G.}~\bibnamefont{Mihaly}},
  \bibinfo{author}{\bibfnamefont{L.}~\bibnamefont{Baldassarre}},
  \bibinfo{author}{\bibfnamefont{A.}~\bibnamefont{Abouelsayed}},
  \bibinfo{author}{\bibfnamefont{C.~A.} \bibnamefont{Kuntscher}},
  \bibinfo{author}{\bibfnamefont{K.}~\bibnamefont{Ohgushi}}, \bibnamefont{and}
  \bibinfo{author}{\bibfnamefont{Y.}~\bibnamefont{Tokura}},
  \bibinfo{journal}{Phys.\ Rev.\ Lett.} \textbf{\bibinfo{volume}{103}},
  \bibinfo{pages}{077205} (\bibinfo{year}{2009}).

\bibitem[{\citenamefont{Tsuda et~al.}(2010)\citenamefont{Tsuda, Morikawa,
  Watanabe, Ohtani, and Arima}}]{tsuda10}
\bibinfo{author}{\bibfnamefont{K.}~\bibnamefont{Tsuda}},
  \bibinfo{author}{\bibfnamefont{D.}~\bibnamefont{Morikawa}},
  \bibinfo{author}{\bibfnamefont{Y.}~\bibnamefont{Watanabe}},
  \bibinfo{author}{\bibfnamefont{S.}~\bibnamefont{Ohtani}}, \bibnamefont{and}
  \bibinfo{author}{\bibfnamefont{T.}~\bibnamefont{Arima}},
  \bibinfo{journal}{Phys.\ Rev.\ B} \textbf{\bibinfo{volume}{81}},
  \bibinfo{pages}{180102(R)} (\bibinfo{year}{2010}).

\bibitem[{\citenamefont{Granroth et~al.}(2006)\citenamefont{Granroth,
  Vandergriff, and Nagler}}]{granroth2006}
\bibinfo{author}{\bibfnamefont{G.~E.} \bibnamefont{Granroth}},
  \bibinfo{author}{\bibfnamefont{D.~H.} \bibnamefont{Vandergriff}},
  \bibnamefont{and} \bibinfo{author}{\bibfnamefont{S.~E.}
  \bibnamefont{Nagler}}, \bibinfo{journal}{Physica B: Condensed Matter}
  \textbf{\bibinfo{volume}{385-86}}, \bibinfo{pages}{1104}
  (\bibinfo{year}{2006}).

\bibitem[{\citenamefont{Granroth et~al.}(2010)\citenamefont{Granroth,
  Kolesnikov, Sherline, Clancy, Ross, Ruff, Gaulin, and Nagler}}]{granroth2010}
\bibinfo{author}{\bibfnamefont{G.~E.} \bibnamefont{Granroth}},
  \bibinfo{author}{\bibfnamefont{A.~I.} \bibnamefont{Kolesnikov}},
  \bibinfo{author}{\bibfnamefont{T.~E.} \bibnamefont{Sherline}},
  \bibinfo{author}{\bibfnamefont{J.~P.} \bibnamefont{Clancy}},
  \bibinfo{author}{\bibfnamefont{K.~A.} \bibnamefont{Ross}},
  \bibinfo{author}{\bibfnamefont{J.~P.~C.} \bibnamefont{Ruff}},
  \bibinfo{author}{\bibfnamefont{B.~D.} \bibnamefont{Gaulin}},
  \bibnamefont{and} \bibinfo{author}{\bibfnamefont{S.~E.}
  \bibnamefont{Nagler}}, \bibinfo{journal}{Journal of Physics: Conference
  Series} \textbf{\bibinfo{volume}{251}}, \bibinfo{pages}{012058}
  (\bibinfo{year}{2010}).

\bibitem[{\citenamefont{Taylor et~al.}(2012)\citenamefont{Taylor, Arnold,
  Bilheaux, Buts, Campbell, Doucet, Draper, Fowler, Gigg, Lynch
  et~al.}}]{taylor2012mantid}
\bibinfo{author}{\bibfnamefont{J.}~\bibnamefont{Taylor}},
  \bibinfo{author}{\bibfnamefont{O.}~\bibnamefont{Arnold}},
  \bibinfo{author}{\bibfnamefont{J.}~\bibnamefont{Bilheaux}},
  \bibinfo{author}{\bibfnamefont{A.}~\bibnamefont{Buts}},
  \bibinfo{author}{\bibfnamefont{S.}~\bibnamefont{Campbell}},
  \bibinfo{author}{\bibfnamefont{M.}~\bibnamefont{Doucet}},
  \bibinfo{author}{\bibfnamefont{N.}~\bibnamefont{Draper}},
  \bibinfo{author}{\bibfnamefont{R.}~\bibnamefont{Fowler}},
  \bibinfo{author}{\bibfnamefont{M.}~\bibnamefont{Gigg}},
  \bibinfo{author}{\bibfnamefont{V.}~\bibnamefont{Lynch}},
  \bibnamefont{et~al.}, \bibinfo{journal}{Bulletin of the American Physical
  Society} \textbf{\bibinfo{volume}{57}} (\bibinfo{year}{2012}).

\bibitem[{hor()}]{horace}
\urlprefix\url{http://horace.isis.rl.ac.uk/}.

\bibitem[{\citenamefont{Buyers et~al.}(1971)\citenamefont{Buyers, Holden,
  Svensson, Cowley, and Hutchings}}]{buyers1971}
\bibinfo{author}{\bibfnamefont{W.~J.~L.} \bibnamefont{Buyers}},
  \bibinfo{author}{\bibfnamefont{T.~M.} \bibnamefont{Holden}},
  \bibinfo{author}{\bibfnamefont{E.~C.} \bibnamefont{Svensson}},
  \bibinfo{author}{\bibfnamefont{R.~A.} \bibnamefont{Cowley}},
  \bibnamefont{and} \bibinfo{author}{\bibfnamefont{M.~T.}
  \bibnamefont{Hutchings}}, \bibinfo{journal}{J. Phys. C: Solid St. Phys.}
  \textbf{\bibinfo{volume}{4}}, \bibinfo{pages}{2139} (\bibinfo{year}{1971}).

\bibitem[{SM()}]{SM_reference}
See Supplementary Material at [URL will be inserted by publisher] for an addition plot of spin-wave curves associated with the parameters in Table~1.


\bibitem[{\citenamefont{Shirane}(1959)}]{shirane59}
\bibinfo{author}{\bibfnamefont{G.}~\bibnamefont{Shirane}},
  \bibinfo{journal}{Acta Cryst.} \textbf{\bibinfo{volume}{12}},
  \bibinfo{pages}{282} (\bibinfo{year}{1959}).

\bibitem[{\citenamefont{Magee}(2010)}]{magee_thesis}
\bibinfo{author}{\bibfnamefont{A.~J.} \bibnamefont{Magee}}, Ph.D. thesis,
  \bibinfo{school}{Royal Holloway}, \bibinfo{address}{University of London}
  (\bibinfo{year}{2010}).



\end{thebibliography}

\pagebreak
\section{Supplementary Material}
\
\begin{figure}[t]
\begin{center}
\includegraphics[width=\columnwidth]{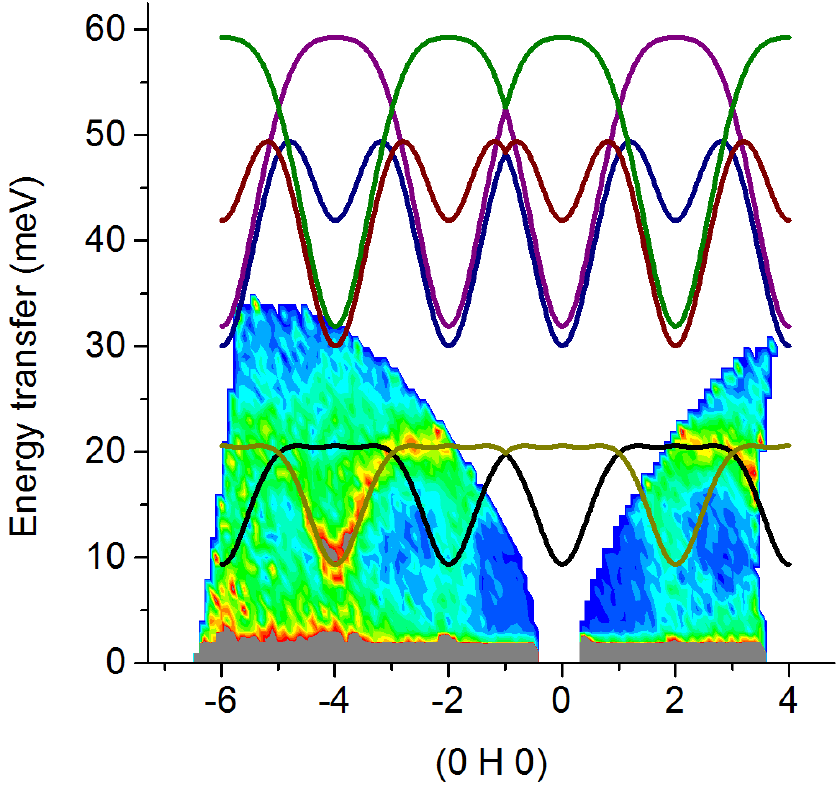}
\end{center}
\end{figure}

\end{document}